\documentclass{emulateapj}

\slugcomment{To appear in Ap.J. Letters}

\shorttitle{Hard X-rays and Gamma-Rays from UHE Protons in Cluster Shocks}
\shortauthors{Inoue, Aharonian \& Sugiyama}

\begin{document}

\title{Hard X-ray and Gamma-ray Emission Induced by Ultra-High Energy Protons
       in Cluster Accretion Shocks}


\author{Susumu Inoue\altaffilmark{1,2,3}, Felix A. Aharonian\altaffilmark{1} and Naoshi Sugiyama\altaffilmark{2}}
\altaffiltext{1}{Max-Planck-Institut f\"ur Kernphysik,
       Postfach 103980, 69029 Heidelberg, Germany}
\altaffiltext{2}{Division of Theoretical Astronomy,
       National Astronomical Observatory of Japan,
       Mitaka, Tokyo, Japan 181-8588}
\altaffiltext{3}{inoue@mpi-hd.mpg.de}

\begin{abstract}
All sufficiently massive clusters of galaxies are expected to be surrounded by strong accretion shocks,
where protons can be accelerated to $\sim 10^{18}$-$10^{19}$ eV under plausible conditions. 
Such protons interact with the cosmic microwave background
and efficiently produce very high energy electron-positron pairs,
which then radiate synchrotron and inverse Compton emission,
peaking respectively at hard X-ray and TeV gamma-ray energies.
Characterized by hard spectra (photon indices $\sim 1.5$)
and spatial distribution tracing the accretion shock,
these can dominate over other nonthermal components
depending on the shock magnetic field.
HESS and other Cerenkov telescopes may detect the TeV emission
from nearby clusters, notwithstanding its extended nature.
The hard X-rays may be observable by future imaging facilities such as NeXT,
and possibly also by ASTRO-E2/HXD.
Such detections will not only provide a clear signature of ultra-high energy proton acceleration,
but also an important probe of the accretion shock itself
as well as magnetic fields in the outermost regions of clusters.
\end{abstract}


\section{Introduction}
\label{sec:intro}

The standard cosmological picture of hierarchical structure formation dictates that
virialized objects such as clusters of galaxies
continue to grow by accreting dark matter and baryonic gas.
It is generically expected that
strong accretion shocks occur around the virial radii of all sufficiently massive clusters,
where infalling gas of relatively low temperature is heated up to the virial value (e.g. Voit 2004).
Here ``accretion'' signifies not only infall of diffuse gas but also minor mergers,
i.e. merging of virialized objects with relatively small masses compared to the cluster.
Such shocks are potentially interesting sites of high energy particle acceleration and nonthermal emission,
possessing high Mach numbers and capable of generating particles with hard spectra at high efficiency
(Ryu et al. 2003; Gabici \& Blasi 2003; Inoue \& Nagashima 2005).
This is in contrast to major merger shocks,
which are likely to play a lesser role for high energy phenomena
on account of their low Mach numbers
(c.f. Totani \& Kitayama 2000; Takizawa \& Naito 2000; Berrington \& Dermer 2003).

Primary electrons, i.e. those accelerated directly by the shocks,
give rise to radio synchrotron emission
as well as inverse Compton (IC) emission by upscattering the cosmic microwave background (CMB)
to UV, hard X-ray and gamma-ray energies
(e.g. Hwang 1997; En\ss lin \& Biermann 1998; Sarazin 1999; Atoyan \& V\"olk 2000;
Waxman \& Loeb 2000; Petrosian 2001).
Such emission should spatially trace the morphology of their injection sites,
which are shell-like regions of Mpc radii for accretion shocks (e.g. Miniati 2003).
On the other hand,
protons interact with the intracluster gas and
generate $\pi^0$ decay gamma-ray emission
(V\"olk, Aharonian \& Breitschwerdt 1996; Berezinsky, Blasi \& Ptuskin 1997;
Pfrommer \& En\ss lin 2004),
as well as synchrotron and IC emission by secondary electrons and positrons from $\pi^{\pm}$ decay
(Blasi \& Colafrancesco 1999),
all of which should follow the centrally-concentrated spatial distribution of thermal gas.

The current observational evidence for such kinds of nonthermal emission is still ambiguous.
Various types of diffuse radio emission in clusters are known
(Kempner et al. 2003; Feretti, Burinaga \& En\ss lin 2004);
however, except for a few cases, they cannot be easily explained
as due to primary electrons nor secondary electrons from p-p interactions (Brunetti 2004).
More controversially, diffuse excess emission over the known thermal component
has been reported for a few clusters in the extreme UV band (e.g. Bowyer et al. 2004)
and in hard X-rays (e.g. Rephaeli \& Gruber 2002; Fusco-Femiano et al. 2004; Henriksen \& Hudson 2004),
which has been interpreted as IC emission from either primary or secondary electrons.
There is no clear case yet for gamma-ray emission from any cluster (Reimer et al. 2003).

Cluster accretion shocks have also been proposed as candidate sources
of ultra-high energy (UHE) cosmic rays (Norman, Melrose \& Achterberg 1995; Kang, Rachen \& Biermann 1997).
However, despite being viable on energetic grounds,
realistic evaluations of the maximum proton energy
seem to fall short of $10^{20}$ eV by 1 to 2 orders of magnitude
(Ostrowski \& Siemieniec-Ozieblo 2000; Jones 2004).

In this letter,
we point out that protons plausibly accelerated to $10^{18}$-$10^{19}$ eV in cluster accretion shocks
should induce important radiative signatures that have not been investigated so far:
synchrotron and IC emission from secondary electron-positron pairs
produced in p-$\gamma$ interactions with the CMB.
Aharonian (2002) and Rordorf, Grasso \& Dolag (2004) have discussed these processes
by assuming that some source, e.g. a powerful radio galaxy,
resides inside a cluster and injects UHE protons into its surroundings.
Here such mechanisms are studied in the context of protons originating in accretion shocks
that may be generic to all massive clusters.
We adopt the WMAP cosmological parameters (Spergel et al. 2003):
$\Omega_m=0.27$, $\Omega_\Lambda=0.73$, $h=0.71$ and $\Omega_b=0.044$.

\section{Model}
\label{sec:model}

For concreteness, we consider a fiducial cluster of mass $M=2 \times 10^{15} M_\sun$
located at distance $D=100$ Mpc, similar to the Coma cluster.
Spherical symmetry and steady state conditions are assumed for simplicity.
The accretion shock radius $R_s$ is taken to be the virial radius, $R_s \simeq 3.2$ Mpc (Voit 2004).
Since gas falls in at a velocity equal to the circular velocity $V_c = (GM/R_s)^{1/2} \simeq 1600$ km/s
and comes to rest downstream of the accretion shock as the shock expands outwards,
the shock velocity relative to the upstream gas is $V_s=(4/3)V_c \simeq 2200$ km/s.
The magnetic field at the shock $B_s$ is uncertain,
as Faraday rotation measurements do not give strong constraints at radii $\ga 1$ Mpc
(Clarke, Kronberg \& B\"ohringer 2001, Johnston-Hollitt \& Ekers 2004).
Some observations of diffuse synchrotron emission
in intergalactic regions outside but near clusters 
suggest field strengths $\sim 0.1$-$1 \mu$G (Kronberg 2004).
We thus choose $B_s$ to be a parameter with $0.1$-$1 \mu$G as a reasonable range,
corresponding to 0.05-5 \% of the gas thermal energy at $R_s$ for our fiducial cluster,
when an isothermal density profile and gas fraction $f_g=\Omega_b/\Omega_m=0.16$ are assumed.

We first evaluate the maximum accelerated proton energy as in Kang et al. (1997).
The timescale $t_{acc}$ for accelerating a proton up to energy $E_p$
in strong shocks is (Blandford \& Eichler 1987)
 \begin{eqnarray}
 t_{acc} &=& {20 \over 3} {\eta r_g c \over V_s^2} \nonumber \\
    &\simeq& 4.4 \times 10^8 {\rm yr} \ \eta {E_p \over 10^{18} {\rm eV}} \left({B_s \over 1 \mu{\rm G}}\right)^{-1} \left({V_s \over 2200 {\rm km/s}}\right)^{-2}
 \label{eq:tacc}
 \end{eqnarray}
where $\eta$ characterizes the mean free path for scattering by magnetic disturbances,
normalized by the particle gyroradius $r_g=E_p/e B_s$.
Observations of nonthermal emission from supernova remnant (SNR) shocks
indicate that $\eta$ is close to unity, i.e. the Bohm diffusion limit
(e.g. V\"olk, Berezhko \& Tsenofontov 2005).
Although it is not guaranteed that this condition is also realized in cluster accretion shocks,
we will assume $\eta=1$ in what follows.

On the timescale of Eq.(\ref{eq:tacc}), 
energy losses by photopair and photopion interactions with the CMB become significant.
We use simple fitting formulae for the timescales of each,
which agree with more accurate calculations (e.g. Yoshida \& Teshima 1993) to within $\sim 20$ \%
in the pertinent energy ranges:
$t_{p\gamma}=\max[t_0 \exp(E_c/E_p), t_1]$,
with $t_0=3.2 \times 10^9$ yr, $t_1=4.6 \times 10^9$ yr and $E_c=3 \times 10^{18}$ eV
for photopair losses,
and $t_0=2.65 \times 10^7$ yr, $t_1=4.5 \times 10^7$ yr and $E_c=2.65 \times 10^{20}$ eV
for photopion losses (see also Waxman 1995).
In addition, we consider the dynamical timescale $t_{dyn}=R_S/V_c \simeq 1.9$ Gyr
to be the lifetime of the accretion shock, limiting the entire process.
On timescales much longer than this, the conditions may change so appreciably
as to invalidate our steady state approximation
(e.g. the cluster may undergo a major merger).
Note that the timescale for adiabatic losses in the expanding accretion shock
will always be larger, $\sim 3 t_{dyn}$, as the shock expansion velocity is $\sim V_c/3$.

The maximum proton energy $E_{\max}$ is obtained by equating $t_{acc}$ and $\min[t_{p\gamma},t_{dyn}]$.
With $B_s=0.1$-$1 \mu$G,
$E_{\max} \simeq 4.5 \times 10^{18} {\rm eV}
\left(B_s/ 1 \mu{\rm G}\right) \left(t_{dyn}/ 1.9 {\rm Gyr}\right)$, always limited by the shock lifetime.
Although this fails to reach the realm of the observed highest energy cosmic rays,
it is noteworthy that $t_{p\gamma}$ is within an order of magnitude of $t_{dyn}$
in the range $E_p \sim 10^{18}$-$10^{19}$ eV.
This implies that during the process, a significant fraction of the proton energy
can be channeled into secondary pairs and consequent radiation. 

The protons are assumed to form an energy distribution
with spectral index 2, appropriate for strong shocks in the test particle limit,
and an exponential cutoff at $E_{\max}$, i.e. $\propto E_p^{-2} \exp (-E_p/E_{\max})$.
This is injected downstream of the shock for a duration $t_{inj}=t_{dyn}$,
at a constant rate normalized by the proton luminosity $L_p$.
We posit $L_p$ to be a fraction $f_p=0.1$ of $L_{acc}$,
the kinetic energy flux through strong accretion shocks,
which can be expressed
 \begin{eqnarray}
 L_{acc} &=& f_g G M \dot{M}/R_s = f_g f_{acc} M V_c^3 / R_s \nonumber \\
         &\simeq& 2.9 \times 10^{46} {\rm erg/s} {f_g \over 0.16} {f_{acc} \over 0.1}
         \left(M \over 2 \times 10^{15} M_\sun\right)^{5/3}
 \label{eq:lacc}
 \end{eqnarray}
where $\dot{M}=f_{acc} V_c^3/G$ is the mass accretion rate (White 1994),
and $f_{acc} \simeq 0.1$ is a factor that was estimated by Keshet et al. (2004)
using numerical simulations.
The numerical value above is for our chosen cosmological parameters,
and is consistent with the simulation results of Miniati (2002)
as well as the analytic estimates of Gabici \& Blasi (2004a).

Given the proton spectrum and injection rate, we follow the method described by Aharonian (2002)
to compute secondary particle production from proton interactions and the resultant emission.
The emission region is taken to be a uniform sphere of radius $R_s$ for simplicity,
although this should actually be a shell behind the shock.
The kinetic equations for the energy distributions of protons and secondary pairs
are solved self-consistently with the emitted spectra,
including the processes of p-$\gamma$ pair production, p-$\gamma$ pion production,
synchrotron and IC radiation from all pairs, and proton synchrotron radiation.
Gamma-ray and pair production from p-p interactions can also be calculated
for a given ambient gas density; we take this to be $n=10^{-6} {\rm cm^{-3}}$,
appropriate for the low density regions around $R_s$.
The diffusion coefficient for protons inside the emission region
is considered to be similar to that at the shock,
implying that diffusive escape is unimportant up to energies 
$E_{esc} \simeq 10^{19} {\rm eV} \ \eta^{-1} (B_s/\mu{\rm G}) (R_s/{\rm 3 Mpc})^2 (t_{inj}/2 {\rm Gyr})^{-1}$.
Intergalactic gamma-ray absorption is incorporated
using the diffuse extragalactic infrared background (IRB) model discussed in Aharonian (2001).

\section{Results}
\label{sec:res}

\begin{figure}
\epsscale{1.0}
\plotone{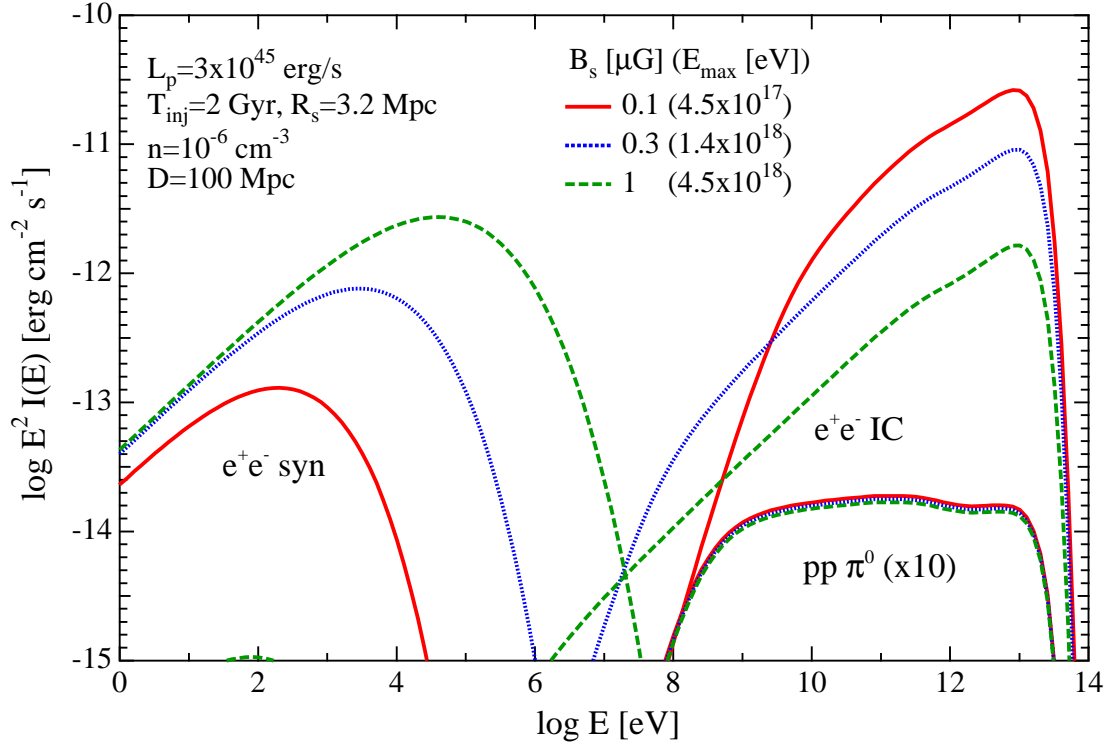}
\caption{Spectra of proton-induced emission from our fiducial cluster accretion shock,
for $B_s=0.1$, $0.3$ and $1 \mu$G. The p-p $\pi^0$ component has been multiplied by 10.}
\label{fig:flux}
\end{figure}

Fig.\ref{fig:flux} shows the broadband spectra of emission induced by protons
in the accretion shock of our fiducial cluster,
for the cases of $B_s=0.1$, $0.3$ and $1 \mu$G.
These respectively imply 
$E_{\max}=4.5 \times 10^{17}$, $1.4 \times 10^{18}$ and $4.5 \times 10^{18}$ eV,
close to or above the CMB photopair threshold but below the photopion threshold. 
The pairs are injected at very high energies $\sim (2m_e/m_p) E_p \sim 10^{15}$ eV
and then rapidly lose energy via synchrotron and IC cooling,
developing an energy distribution $\propto E_e^{-2}$
and emitting hard spectra with photon indices $\Gamma \sim 1.5$ (Aharonian 2002).
Spectral peaks occur at $\sim 1$-$100$ keV for synchrotron and at $\sim 10$-$100$ TeV for IC,
while intergalactic absorption by the IRB
imposes a further cutoff above $\sim 10$ TeV in the observed flux.
Note the large influence of $B_s$,
which affects both the synchrotron cooling rate and the peak energy
as well as the amount of pair injection by changing
the fraction of protons above threshold through $E_{\max}$.
Lower $B_s$ causes lower $E_{\max}$ and less pair injection
but also reduces the synchrotron cooling,
leading overall to increased IC and decreased synchrotron luminosities;
for higher $B_s$, vice-versa.
The relative luminosities together with the peak frequency locations should allow
$B_s$ to be reliably determined from observations.
For our range of $B_s$,
the emitted power is mostly dominated by IC gamma-rays and can be as large as $\sim 10^{44}$ erg/s,
despite Klein-Nishina suppression being effective near the peak.
This testifies that protons above threshold are radiatively efficient
in the sense that a major portion of their energy is converted to radiative pairs.
With $n = 10^{-6} {\rm cm^{-3}}$,
$\pi^0$ gamma-rays and secondary pair radiation from p-p interactions
only make a miniscule contribution,
although they should be more pronounced from the central, higher density regions of the cluster,
which is not considered here.

Owing to the hard spectra and high radiative efficiency from protons,
the p-$\gamma$ pair (hereafter PGP) emission
may dominate the hard X-ray and TeV gamma-ray bands
over other components from primary electrons or p-p interactions,
which should have spectra with $\Gamma \sim$ 2 or steeper at these energies.
Conversely, where PGP emission is not strong, as in the GeV range,
the latter processes may still prevail.
For example, if the electron injection efficiency
relative to the accretion kinetic energy is $f_e=0.01$, the primary IC luminosity
$L_{IC} \simeq (2 \ln \gamma_{e,\max})^{-1} f_e L_{acc} \simeq 8 \times 10^{42}$ erg/s for our fiducial cluster,
where $\gamma_{e,\max} \sim 3 \times 10^7$ is the maximum electron Lorentz factor
determined from the balance between acceleration and cooling times.
This leads to flux $\sim 7 \times 10^{-12} {\rm \ erg \ cm^{-2} \ s^{-1}}$ from $D=100$ Mpc,
close to the $\sim 100$ MeV EGRET upper limit for Coma (Reimer et al. 2003);
nevertheless, PGP emission can be dominant at TeV if $B_s \la 0.3 \mu$G.
Note that although $f_e$ is quite uncertain and values up to $\sim$ 0.01 may be allowed,
it may in fact be significantly lower.
The spatial distribution of PGP emission should be similar to primary IC,
tracking the projected ring-like morphology of the accretion shock,
so that p-p $\pi^0$ gamma-rays may still be prominent for the central core regions
in spatially resolved maps.

To assess the detectability of PGP emission with various observational facilities,
we note that even though the full angular extent of the accretion shock diameter can be as large as
$3.6^\circ (R_s/3.2 {\rm Mpc}) (D/100 {\rm Mpc})^{-1}$,
the expected ring-like distribution is highly nonuniform;
we thus choose to compare model fluxes with instrumental sensitivities for
sources uniformly extended by $1^\circ$.
Most promising for detection is the PGP IC component in TeV gamma-rays,
where the currently operating HESS\footnote{\url{http://www.mpi-hd.mpg.de/hfm/HESS/HESS.html}}
array of Cerenkov telescopes
can achieve point source sensitivities
$\sim 10^{-13} {\rm erg \ cm^{-2} \ s^{-1}}$ at 1 TeV in $\sim$100 hr exposure time.
With its $0.1^\circ$ angular resolution, the $1^\circ$ extended source sensitivity 
should be roughly $10^{-12} {\rm erg \ cm^{-2} \ s^{-1}}$,
so that even our least luminous case of $B_s=1\mu$G may be detectable,
not to mention the more luminous, lower $B_s$ cases.
Furthermore, its $5^\circ$ field of view may allow clear imaging
of the annular morphology with a single pointing.
Detecting the weaker PGP IC emission at GeV energies may be difficult
even for the GLAST\footnote{\url{http://www-glast.stanford.edu/}} mission,
with expected point source sensitivities reaching $\sim 3 \times 10^{-13} {\rm erg \ cm^{-2} \ s^{-1}}$ at 1 GeV
during a 1 yr survey.
Moreover, it could likely be that other emission components such as primary IC 
show up above the PGP emission levels in this band (e.g. Miniati 2003; Gabici \& Blasi 2004a).

The PGP synchrotron component should be interesting for future facilities such as NeXT (Takahashi 2005),
capable of imaging hard X-rays in the 8-80 keV range at $\sim$ 30" resolution.
The projected sensitivity for a $1^\circ$ source is
$\sim 10^{-12} {\rm erg \ cm^{-2} \ s^{-1}}$ at $30$ keV in a 100 ks exposure,
so the cases of $B_s \ga 0.3 \mu$G should be clearly detectable,
and the accretion shock morphology may be studied in fine detail.
The non-imaging HXD instrument on the soon-to-be launched ASTRO-E2 (Takahashi 2005),
with a sensitivity of $\la 3 \times 10^{-12} {\rm erg \ cm^{-2} \ s^{-1}}$ at 20 keV,
as well as instruments onboard INTEGRAL
may also have chances of detection if $B_s \ga 1\mu$G.

\begin{figure}
\epsscale{1.0}
\plotone{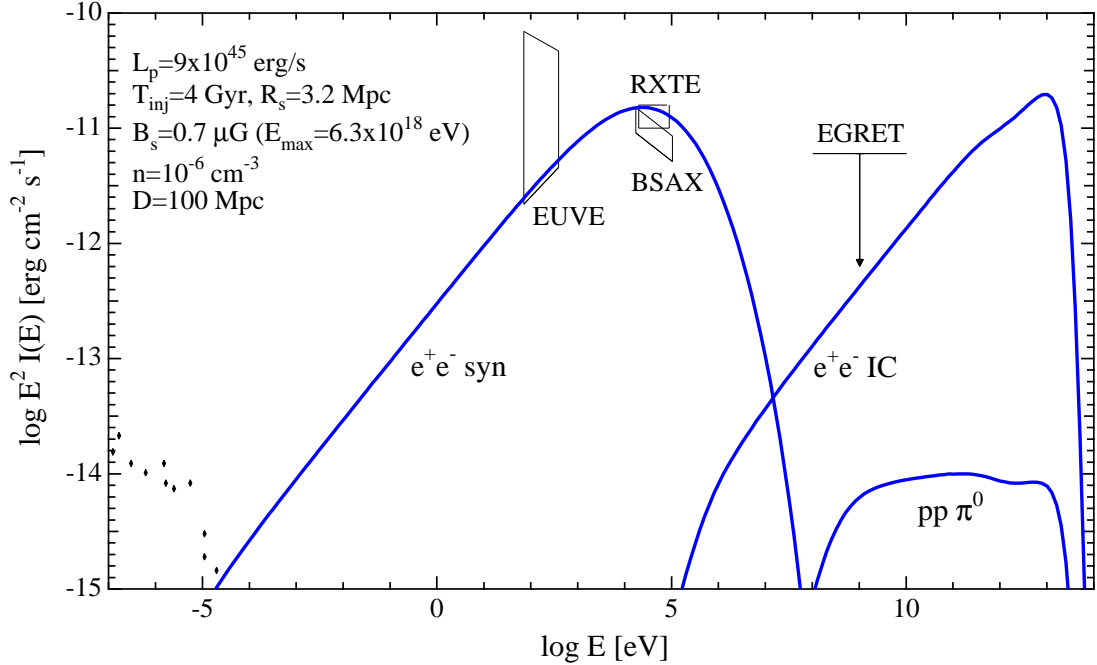}
\caption{Spectra of proton-induced emission for the model parameters labelled in the figure (see also text),
compared with observational data for the Coma cluster from EUVE, BeppoSax, RXTE and EGRET
compiled in Reimer et al. (2004).
}
\label{fig:coma}
\end{figure}

Finally, it is interesting to consider whether the model can explain
the reported excess UV and hard X-ray emission for the Coma cluster.
In Fig.\ref{fig:coma},
multifrequency data compiled in Reimer et al. (2004)
are compared with a model with $B_s=0.7 \mu$G, $L_p=9 \times 10^{45} {\rm erg/s}$ and $T_{inj}=4$ Gyr,
otherwise being the same as in Fig.\ref{fig:flux}.
Despite appearing rather consistent,
we recall that the UV and hard X-ray data correspond only to the inner $\la 1^\circ$ region,
whereas the model emission should be extended out to projected radius $\simeq 1.8^\circ$.
To explain the observations with just the fraction of emission projected onto the core,
more extreme parameters for $L_p$ and/or $t_{dyn}$ may be necessary.
Moreover, Bowyer et al. (2004) argue that the UV emission is spatially correlated
with thermal X-rays, favoring a p-p secondary origin.
In any case, further observations at these as well as gamma-ray energies
should provide conclusive answers.
Note that neither the PGP nor p-p secondary synchrotron component from the accretion shock
can account for the radio emission (Thierbach, Klein \& Wielebinski 2003),
calling for a different origin
such as particles accelerated by turbulence
(Brunetti 2004; Fujita, Takizawa \& Sarazin 2003).

\section{Discussion}
\label{sec:disc}

Successful observations of the PGP emission discussed above
entails a number of important implications, by providing:
1) a clear signature of UHE protons in cluster accretion shocks,
and hence a test of particle acceleration theory on the largest scales;
2) a sensitive probe of magnetic fields in the outermost regions of clusters, which is crucial for
understanding the origin of intergalactic and intracluster magnetic fields,
propagation of UHE cosmic rays,
etc.;
3) a potential probe of the accretion shock itself,
which still lacks firm observational evidence
despite being a robustly predicted cosmological phenomenon;
4) a useful tool for gamma-ray absorption studies of the IRB,
owing to the hard, steady TeV spectra, as opposed to the highly variable spectra of blazars.

Our assumption of $\eta=1$ as inferred for SNRs
may not be completely justified for cluster accretion shocks.
Larger $\eta$ would decrease $E_{\max}$ and hence the amount of pair production and PGP emission;
e.g. for the case of $B_s=0.3 \mu$G, $\eta=3$ and $10$ lead to IC luminosities lower
by factors of $\sim 3$ and $\sim 20$, respectively.
On the other hand, the proton injection efficiency can be higher than our choice of $f_p=0.1$.
Increased emission may also result from
proton spectra harder than 2 due to nonlinear acceleration effects 
(Gabici \& Blasi 2004b; Kang \& Jones 2005),
as well as from tertiary pairs produced in $\gamma$-$\gamma$ interactions
of PGP gamma-rays with the CMB (Timokhin, Aharonian \& Neronov 2004).

The present study may be more appropriate
for clusters that are relatively relaxed
rathen than those associated with a current or recent major merger,
although accretion shocks should be present even in the latter case (Ryu et al. 2003).
The accretion process in reality is neither spherical nor steady,
involving clumpy merging along filamentary structures.
However, as long as the merging leads to strong, high Mach number shocks,
proton acceleration may proceed even through successive merging events (Kang \& Jones 2005).
At any rate, more detailed investigations of the processes discussed here
should be carried out by incorporating them in numerical simulations of large scale structure formation.

\acknowledgments
We acknowledge fruitful discussions with E. Churazov, B. Dingus, T. En\ss lin, 
S. Gabici, J. Kirk, F. Miniati, K. Nakazawa and T. Takahashi.

\end{document}